# Recent progress in the assembly of nanodevices and van der Waals heterostructures by deterministic placement of 2D materials

*Riccardo Frisenda[1,*], Efrén Navarro-Moratalla[2], Patricia Gant[1], David Pérez De Lara[1], Pablo Jarillo-Herrero[2], Roman V. Gorbachev[3] and Andres Castellanos-Gomez[4,*]*

[1] *Instituto Madrileño de Estudios Avanzados en Nanociencia (IMDEA Nanociencia), Madrid, E-28049, Spain.*

[2] *Department of Physics, Massachusetts Institute of Technology, Cambridge, Massachusetts 02139, USA*

[3] *National Graphene Institute, University of Manchester, Manchester M13 9PL, United Kingdom*

[4] *Materials Science Factory, Instituto de Ciencia de los Materiales de Madrid (ICMM-CSIC), Madrid, E-28049, Spain.*

\* riccardo.frisenda@imdea.org and andres.castellanos@csic.es

**ABSTRACT**

Designer heterostructures can now be assembled layer-by-layer with unmatched precision thanks to the recently developed deterministic placement methods to transfer two-dimensional (2D) materials. This possibility constitutes the birth of a very active research field on the so-called van der Waals heterostructures. Moreover, these deterministic placement methods also open the door to fabricate complex devices, which would be otherwise very difficult to achieve by conventional bottom-up nanofabrication approaches, and to fabricate fully-encapsulated devices with exquisite electronic properties. The integration of 2D materials with existing technologies such as photonic and superconducting waveguides and fiber optics is another exciting possibility. Here, we review the state-of-the-art of the deterministic placement methods, describing and comparing the different alternative methods available in the literature and we illustrate their potential to fabricate van der Waals heterostructures, to integrate 2D materials into complex devices and to fabricate artificial bilayer structures where the layers present a user-defined rotational twisting angle.

1. **Introduction:**

Richard P. Feynman said in his seminal lecture 'There is plenty of room at the bottom': "… What could we do with layered structures with just the right layers? What would the properties of materials be if we could really arrange the atoms the way we want them? They would be very interesting to investigate theoretically. I can't see exactly what would happen, but I can hardly doubt that when we have some control of the arrangement of things on a small scale we will get an enormously greater range of possible properties that substances can have, and of different things that we can do…".[1] In this lecture, considered as the birth of the field of nanoscience, Feynman envisioned a future where artificial materials could be engineered with on demand properties. The isolation of 2D materials has brought this vision one step closer to reality. In fact, during his Nobel lecture Konstantin Novoselov [2] reminded us that van der Waals heterostructures,[3–5] fabricated by stacking individual layers of different 2D materials, constitutes an experimental realization of these artificial materials proposed by Feynman.

Since 2010, the research field of van der Waals heterostructures has grown exponentially spurred by the continuous development of ingenious experimental techniques that allow atomically thin materials to be placed on the desired locations with an unprecedented degree of control and accuracy.[6–10] These deterministic placement methods have also opened the door to observe new physical phenomena[11–13] and to assemble complex devices [14–17] that would be otherwise impossible to fabricate by conventional bottom-up approaches.

One of the first important realizations in the field of van der Waals artificial structures was the fabrication of superlattices with arbitrary periodicity in a simple and effective way, making possible the exploration of the moiré patterns that emerge in twisted artificial bilayers (graphene on graphene or graphene on boron nitride). In fact, the deterministic transfer methods represent the easiest way to produce layered materials where incommensurate layers are stacked at will. A different context where the deterministic transfer methods have proven their efficacy is in the capping of 2D materials to improve their electronic properties. For example, graphene and $MoS_2$ transistors displayed a large increase in their mobility when capped between two hexagonal boron nitride flakes. Moreover, the deterministic transfer of boron nitride flakes can also be used as a protection layer against external agents which might be crucial for applications using unstable 2D materials as it will be discussed later.

The aim of this Tutorial Review is to provide a comprehensive description of the recently developed deterministic transfer methods of 2D materials providing a good starting point for researchers that are beginning to work on the field of 2D materials and van der Waals heterostructures. We first introduce the general experimental requirements for a deterministic placement setup, discussing a few real examples from the authors' laboratories. Following on from this, an in-depth description and a critical comparison between the different methods is proposed, allowing one to identify the optimal method for each application. Finally, we discuss some examples that illustrate the potential of the deterministic transfer methods in the fabrication of van der Waals heterostructures and in the fabrication of layer-by-layer assembled electronic and optoelectronic devices.

## 2. Deterministic placement methods

In this section, we will present a systematic classification of the different deterministic transfer methods reported in the literature, allowing for a direct comparison between them.

Despite the differences between these methods they all rely on very similar experimental tools: namely a long working distance optical inspection system in combination with one or two micropositioning systems. Figure 1a shows an image of a typical deterministic placement setup based on a modified optical microscope equipped with long working distance objectives and two manually actuated micropositioners (one to move the acceptor surface and another one to move the flake to be transferred). Other systems (see Figure 1b) are based on zoom lenses instead of on modified optical microscopes. These systems are less expensive and have the advantage of longer working distance at a cost of a slight reduction in the image quality. Figure 1c also shows how the deterministic transfer setups can be implemented inside a glovebox to achieve control over the environment during the transfer. In the Challenges and Outlook section, we will discuss the relevance of the development of these setups with controlled environment to reduce the presence of interlayer adsorbates and to work with materials that degrade upon exposure to air.

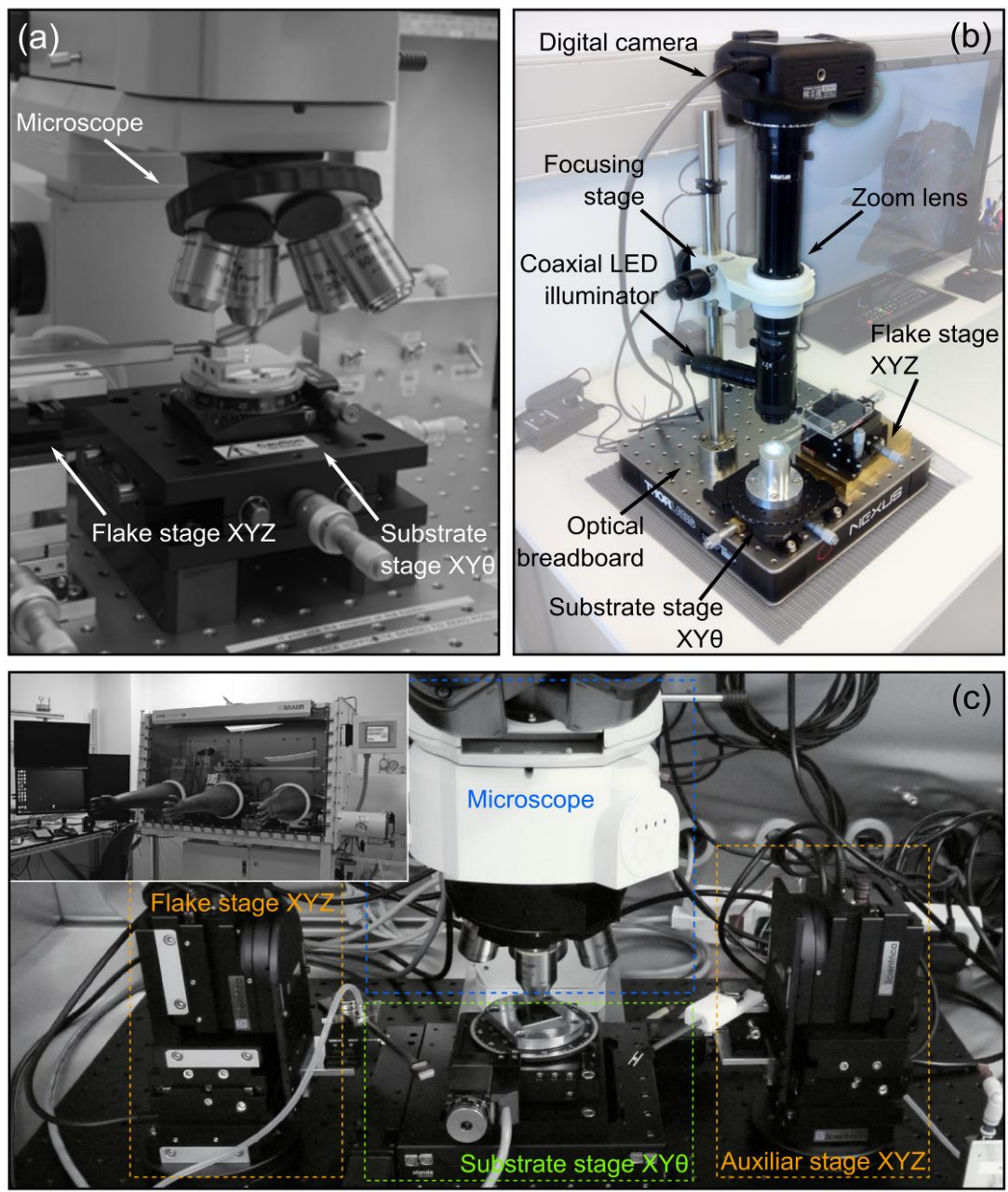

**Figure 1**. **Deterministic transfer setups.** Images of different deterministic placement experimental setups. (a) Experimental setup based on a modification of a metallurgical optical microscope equipped with long working distance objectives. (b) Setup implemented using a zoom lens with coaxial illumination. (c) Deterministic placement setup mounted inside a glovebox to ensure control of the environmental conditions during the transfer process. The stages are fully motorized and they can be controlled with joysticks from outside the glovebox. Panels (a) and (c) are pictures of the transfer setup at Manchester; panel (b) is a picture of the setup at Madrid.

### 2.1. PMMA carrying layer method

The PMMA (polymethyl methacrylate) carrying layer transfer method was pioneered by Dean *et al.* to fabricate high mobility graphene devices on top of mechanically exfoliated hexagonal boron nitride flakes instead of the standard $SiO_2$ substrates.[6]

In the PMMA carrying layer method, illustrated in Figure 2, the desired flake is exfoliated onto a Si/$SiO_2$ carrier substrate coated with a polymer stack consisting of a water-soluble layer (Mitsubishi Rayon aquaSAVE®) and PMMA (**1**). By tuning the thickness of the PMMA one can reach an ideal optical contrast of the flake, useful for optical identification. The substrate is then immersed in deionized water (**2**) where the soluble polymer can dissolve, leaving the hydrophobic PMMA layer carrying the flake floating on top of the water surface (**3**). The PMMA layer can then be attached to a glass transfer slide clamped to a micromanipulator and positioned on an optical microscope (**4**). Using the microscope and the micromanipulator one can locate the flake and align it precisely with the target substrate (which can contain markers, a different flake, pre-patterned electrodes…). The flake and the substrate can be brought into contact (**5**) and then the glass slide can be raised slowly allowing a gentle detaching of the flake from the PMMA (**6**). A different way of detaching the flake from PMMA is to dissolve the PMMA layer in acetone after baking the stack at ≈100 °C. We have based the description of the deterministic transfer method on Ref. [6] but there are variations over the original transfer method developed by Dean *et al*. [18–22]. For example, in Refs. [18,19,22] a water soluble layer of polyvinyl alcohol (PVA) is used instead of the aquaSAVE. In Ref. [21] the authors didn't use any polymer sacrificial layer (neither aquaSAVE nor PVA) and they delaminate the PMMA carrier layer from the substrate through a partial etching of the $SiO_2$ layer in a KOH solution. In Ref.[23] Taychatanapat *et al.* demonstrated a PMMA carrying layer transfer method where the polymer stack acting as carrier layer (PMMA/PVA) is mechanically peeled off from the substrate with an adhesive tape, avoiding all the contact of the sample with water and thus being an all-dry transfer method which advantages of the absence of capillary forces during the transfer.

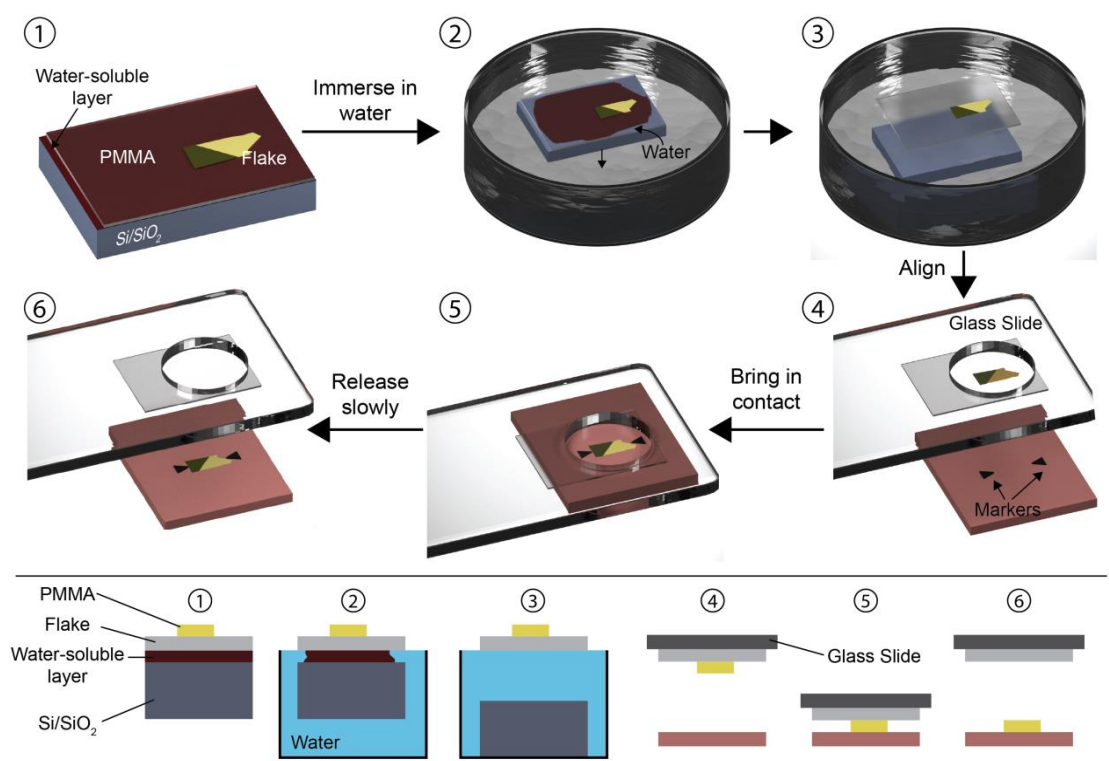

**Figure 2**. **The PMMA carrying layer transfer method.** The designated flake to be transferred is exfoliated onto a Si/SiO$_2$ substrate which has been coated with a water-soluble polymer layer and PMMA (1). The stack is then immersed in water (2) where the water-soluble layer can dissolve leaving the PMMA layer carrying the flake floating on the water surface (3). The PMMA is then attached to a glass slide connected to a micromanipulator with the flake facing down (4). With the help of a microscope the flake is aligned with the target substrate and is then brought in contact (5). By gently separating the PMMA from the final substrate the flake get transferred.

### 2.2. Elvacite sacrificial layer

Zomer *et al*. developed an alternative sacrificial layer based transfer method by employing a polymer with a low glass transition temperature (Elvacite®) instead of PMMA.[8] The Elvacite sacrificial layer technique steps are shown in Figure 3. A flexible and transparent adhesive tape (Pritt®) is attached to a glass slide (adhesive side facing the glass slide). Then a layer of methyl/n-butyl methacrylate copolymer (Elvacite 2550 acrylic resin) dissolved in methyl isobutyl ketone (MIBK) is spin-coated (~1 µm thick). The stack is then baked for 10 minutes at 120 °C in order to remove the MIBK solvent from the copolymer and then the 2D material of choice is exfoliated on top of the stack (**1**). The target substrate for the transfer is mounted on a sample holder that can be heated up and the temperature is set in the range between 75ºC and 100°C. Using an optical microscope and a micromanipulator one can align the flake with the target substrate (**2**) and bring the flake in contact with the hot substrate. Once the polymer touches the hot substrate, it melts (**3**) and it is released from the adhesive tape layer by lifting the glass slide (**4**). After the transfer procedure, the polymer can be removed using acetone. Hunt *et al.*[13] presented a modification of the Elvacite sacrificial layer transfer method described above that consists of replacing the Elvacite polymer with a MMA (methyl methacrylate) sacrificial layer that requires slightly higher temperature than the Elvacite to be released.

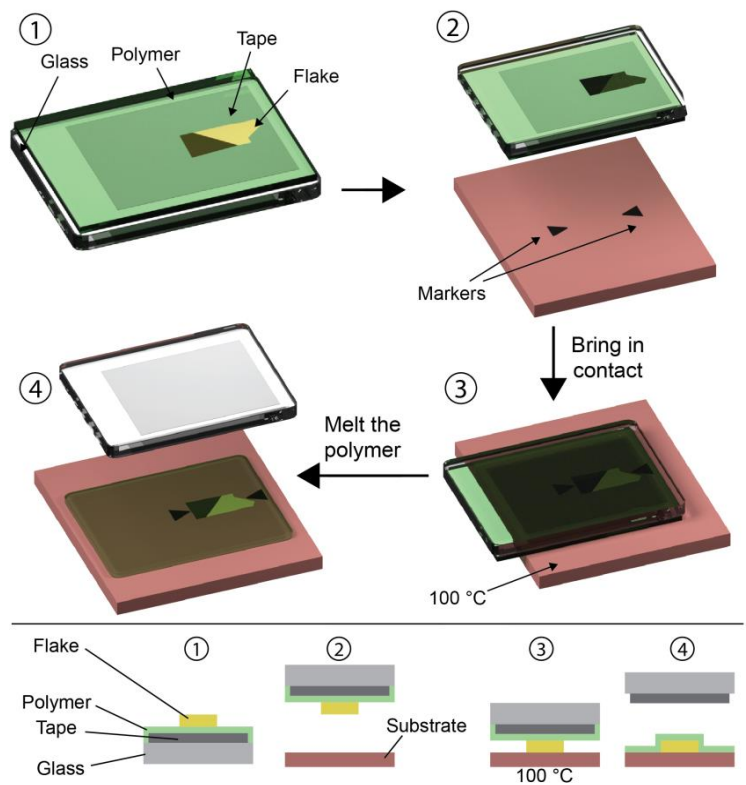

**Figure 3**. **The Elvacite sacrificial layer transfer method.** The target flake is exfoliated onto a layer of Evalcite deposited onto adhesive tape attached to a glass slide (1). The stack is then positioned onto a target substrate and the flake is aligned with pre-existing features of the substrate (2). The substrate is heated up at 100 °C and the stack can then be brought in contact with the hot substrate (3) which melts the Evalcite layer leaving the capped flake on the final substrate (4).

### 2.3. Wedging transfer method

Schneider *et al*. developed a water-based method for transferring 2D materials onto surfaces of various shapes and compositions.[7] The transfer occurs through the intercalation of a layer of water between a hydrophilic substrate and a hydrophobic polymer thin film where the 2D materials are locked.

The wedging transfer technique is schematically depicted in Figure 4. A hydrophilic substrate (*e.g.* $SiO_2$), with the target flake exfoliated on top (**1**), is dipped into a solution of a hydrophobic polymer (cellulose acetate butyrate, 30 mg/mL in ethyl acetate, Sigma-Aldrich) (**2**). The substrate is immersed in water (**3**) and a layer of water is intercalated (wedged) between the hydrophilic $SiO_2$ and the hydrophobic polymer carrying the flake delaminating the polymer film that remains floating at the water/air interface. The transfer of the flake is achieved by pumping down the water (**4**), which lowers the water level and brings the polymer in contact with a target substrate lying below (**5**). To align the flake with the target substrate, a needle coupled to a micromanipulator can be used to gently position the floating polymer at will. Once the flake is deposited onto the target substrate, the polymer layer is removed by dissolving it into a solvent (in this case ethyl acetate) (**6**). In Ref. [22] the cellulose acetate butyrate polymer layer is substituted for PMMA which is delaminated from the substrate by partial etching of the $SiO_2$ layer with NaOH or by water intercalation.

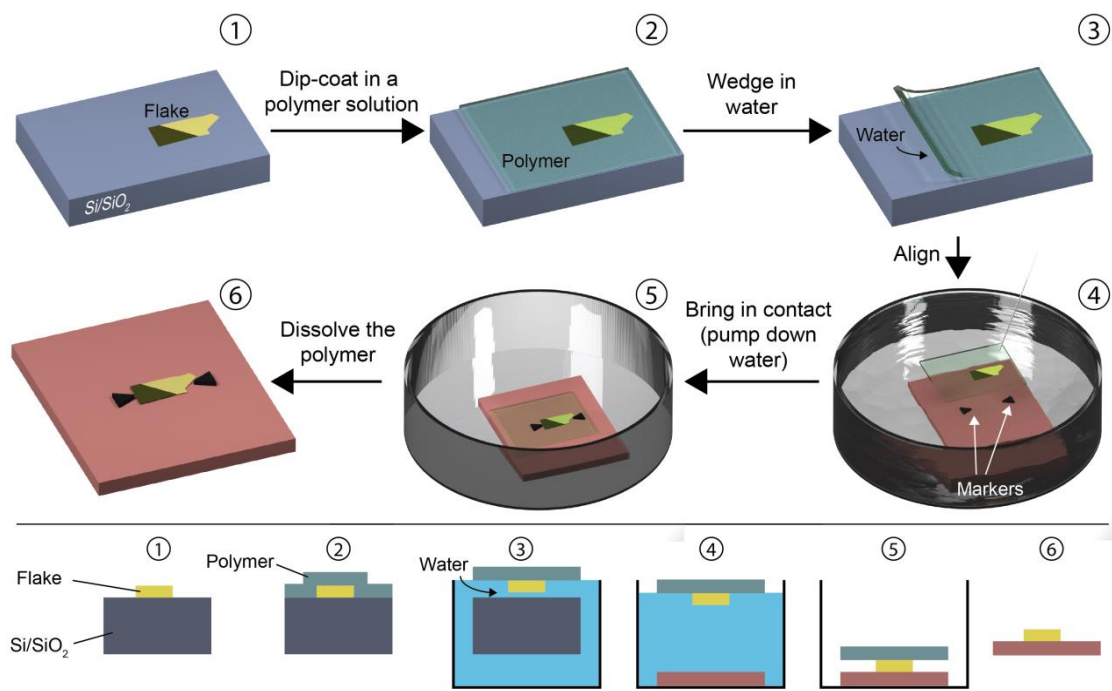

**Figure 4**. **The wedging transfer method.** The flake to be transferred is exfoliated onto a hydrophilic substrate such as Si/$SiO_2$ (1) and then covered by a layer of hydrophobic polymer (2). The stack is then immersed in water where the water molecules can intercalate between the $SiO_2$ and the polymer carrying the flake (3). The polymer film carrying the flake remains floating on the water surface where it can be aligned with the help of a needle to the final substrate

(4). By pumping down the water the two can be brought in contact (5). Then the polymer can be dissolved leaving the flake transferred onto the final substrate (6).

As the lifting of the flake and its subsequent alignment and transfer is carried out in water, the wedging transfer method has severe limitations to fabricate freely-suspended 2D materials as they may collapse because of the capillary forces. Moreover, the fabricated samples may present trapped water blisters and/or wrinkles produced during the drying process (see Figure 5).[24] On the other hand, the wedging transfer method is probably one of the most versatile methods that allows one to transfer flakes of 2D materials on top of curved and uneven surfaces.

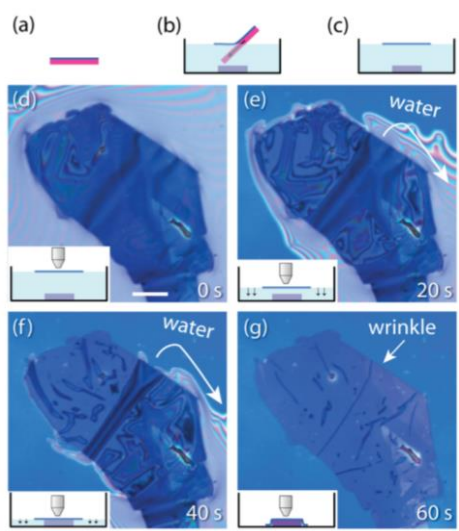

**Figure 5**. **Example of wrinkles and bubbles induced during the wedging transfer.** The polymer layer carrying a few-layer graphene flake is delaminated from a $SiO_2$/Si substrate and floated on water (a) to (c). (d)-(f) Optical pictures of the wet transfer process. Once water is completely drained, there are wrinkles left behind. The scale-bar in (d) is 50 µm. The schematics in the insets show the water level in each panel. This figure has been reproduced from Ref. [24] with permission.

### 2.4. PDMS deterministic transfer method

Unlike the previously described deterministic transfer methods that rely on the use of sacrificial polymer layers and/or wet chemistry to some extent (spin coating/dip coating of the polymer layer, dissolving/wedging, etc). Castellanos-Gomez *et al*. developed an all-dry transfer method that relies on viscoelastic stamps. This method does not employ any wet chemistry step which was found very advantageous in terms of transfer speed and to fabricate devices with freely suspended 2D materials as there are no capillary forces involved in the process.[10]

This technique, shown in Figure 6, is based on the viscoelastic properties of PDMS (polydimethylsiloxane). Initially a flake is exfoliated onto a commercially available viscoelastic PDMS stamp (Gelfilm by Gelpak®) by mechanical exfoliation with Nitto tape (**1**). The PDMS stamp is then attached overhanging in a cantilever configuration to a glass slide connected to a micromanipulator (**2**). By monitoring the process with an optical microscope one can precisely align the flake on the PDMS with the target substrate (**3**) and then bring the two in contact (**4**). In order to release the flake from the PDMS stamp, the glass slide is raised slowly in such a way that the PDMS can detach gently from the substrate (**5**) leaving the flake in the chosen position on the substrate (**6**).

We have based the description of the PDMS based deterministic transfer method on Ref.[10] but there are variations [25,26] over the original transfer method developed by Castellanos-Gomez and co-workers. For example, Ref.[25] describes a transfer method that it is an hybrid between the wedging transfer (section 2.3) and the PDMS dry transfer method described in this section.

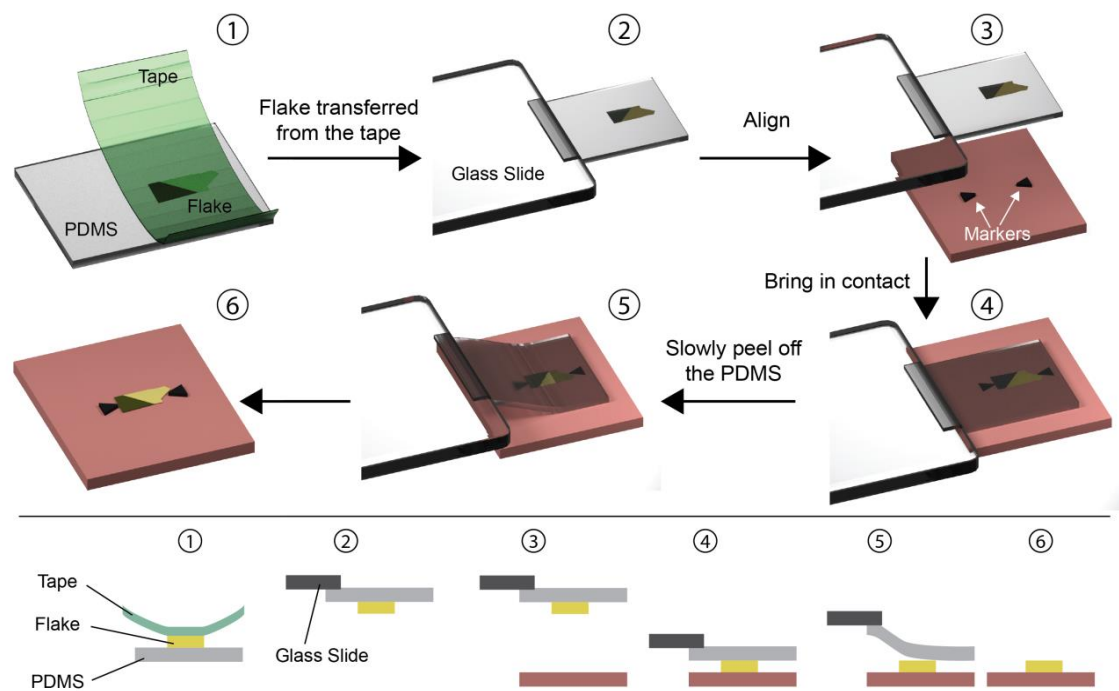

**Figure 6**. **The PDMS dry transfer method.** The flake to be transferred is exfoliated onto a PDMS stamp (1) and the stamp is then attached to a glass slide connected to a micromanipulator (2). Using a microscope the flake can be aligned with the final substrate (3) and brought in contact (4). By slowly peeling of the PDMS stamp (5) the flake can be deposited on the substrate (6).

### 2.5. Van der Waals pick-up transfer methods

Wang *et al.* introduced a new concept of deterministic transfer method where the van der Waals interaction between different 2D materials is exploited to transfer flakes without contacting the 2D materials with any polymer throughout the whole process, reducing impurities trapped between the layers.[9] The van der Waals pick-up transfer methods are based on lifting 2D materials with the help of a hexagonal boron nitride flake to transfer the stack somewhere else (see Figure 7). The target flake to be transferred is exfoliated onto a $SiO_2$ surface. To pick-up the flake and transfer it onto the desired position, one uses a stamp formed by a PDMS film covered with a layer of polypropylene carbonate PPC and with a boron nitride (BN) flake exfoliated onto it. The stamp is mounted in a micromanipulator for control and alignment (**1**). The stamp is lowered until the BN flake contacts the target flake (**2**). As both BN flake and target flake are atomically flat and clean surfaces, their contact area (and thus the adhesion force) can be very large. By slowly lifting up the stamp, the flake can be picked-up from the substrate (**3**) and precisely aligned with the target substrate (**4**). The stacked flakes are then brought in contact with the target substrate (**5**) where they can be transferred by releasing gently the contact (**6**).

We have based the description of the van der Waals pick-up transfer method on Ref.[9] but there are reports of slight variations over that method.[27–29] For instance, in the Supporting Information of Ref.[27] an alternative van der Waals pick-up, where repetitive stacks are transferred without using PDMS as a carrier layer, is described.

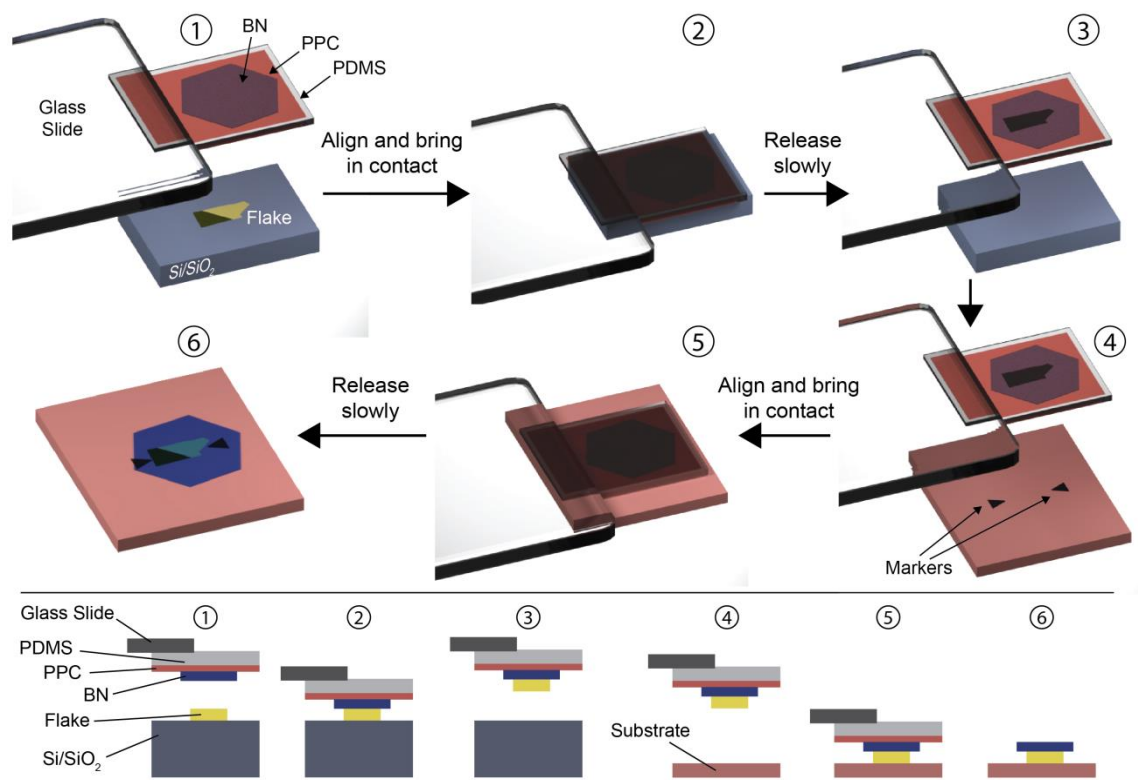

**Figure 7**. **The van der Waals pick-up transfer method.** The flake to be transferred is exfoliated onto a Si/SiO$_2$ substrate (1). A stack composed of PDMS with a layer of PPC and a boron nitride (BN) flake is attached to a glass

slide. The BN is brought in contact with the target flake (2) and by releasing the contact slowly the target flake can be picked up remaining attached to the BN flake (3). The stack can then be positioned on top of the target substrate (4) and then brought in contact (5). By slowly releasing the contact one can transfer the target flake capped by the BN flake (6).

## 2.6. Comparison between the different deterministic transfer methods

Although a quantitative comparison between the different transfer methods is difficult to make, in this section we will provide a qualitative comparison between the different methods described above in terms of some parameters that are relevant in the experiments: how clean are the as-fabricated samples, how easy is to implement the deterministic transfer setup, how long does it take to transfer a flake, etc.

Regarding the cleanness, the Elvacite sacrificial layer and the wedging transfer methods yield samples that have been in direct contact with polymers that have to be dissolved through wet chemistry process. Also, the wedging transfer method is more difficult to implement than the other methods as it requires a pump to drain the water during the transfer but it has the advantage that it allows one to transfer flakes on top of curved and uneven surfaces where other transfer methods may fail. In the PMMA carrier layer method the flakes are in contact with the PMMA layer but they can be transferred without needing to dissolve the PMMA layer. In the PDMS dry transfer method the flakes are also in direct contact with the PDMS carrier layer. The van der Waals pick-up method provides the cleaner samples as the flake to be transferred is never in direct contact with any polymer nonetheless this method only allows for the transfer of heterostructures (such as flakes encapsulated between two boron nitride flakes) and it requires several steps making it very time consuming. In terms of easiness of implementation and speed the PDMS dry transfer method advantages the other deterministic transfer methods as it does not require any wet chemistry for the preparation of the carrier layer (the other methods require at least a step of spin coating or dip coating plus a baking step) as the PDMS layer can be directly purchased (e.g. Gelfilm from Gelpak®) and the flakes are peeled off directly from the PDMS carrier layer (thus it is not necessary to dissolve the carrier layer like in the Elvacite or wedging transfer methods). Table 1 briefly summarizes the comparison discussed above.

| Method | Cleanness | Easiness | Speed | Notes |
| --- | --- | --- | --- | --- |
| PMMA carrier layer | *** | *** | *** | Spin-coating is needed, direct contact with polymer. |
| Elvacite sacrificial layer | * | *** | *** | Capillary forces, spin-coating is needed, direct contact with polymer. |
| Wedging | * | ** | *** | Capillary forces, dip-coating is needed, difficult alignment, direct contact with polymer, transfer over curved or uneven surfaces is possible. |
| PDMS dry transfer | *** | ***** | ***** | Direct contact with polymer. |
| Van der Waals pick-up | ***** | * | ** | Spin-coating is needed, several steps involved, only works to transfer heterostructures, direct contact with the polymer only for the topmost layer. |

**Table 1**. **Comparison between the different deterministic placement methods.** Qualitative comparison in terms of cleanness, easiness and speed between the different deterministic placement methods described in the text. Comments about their main drawbacks are also included in the table.

## 3. Fabrication of van der Waals heterostructures

The layer-by-layer assembly of van der Waals heterostructures is conceptually similar to molecular beam epitaxy (MBE) and, therefore, many heterostructures and devices are drawn from analogies with structures previously made by MBE and studied by a large semiconductor community for three decades. There are, however, principle differences. Firstly, instead of III-V and II-VI semiconductors that are the dominant compounds used in MBE, one can fabricate van der Waals hterostructures employing over a hundred different layered crystals that do not require a particular lattice matching or other specific growth conditions. Secondly, 2D materials obtained by the cleavage of bulk layered crystals are a readily available source of materials with high quality electronic transport properties without going through the process of optimizing the growing conditions like in MBE. Thirdly, atomically-sharp interfaces are often very difficult to obtain through the use of MBE and other epitaxial-growth techniques as interdiffusion generally creates a transitional region between two epitaxially grown materials, which typically can extend over several layers. In contrast, the assembly of van der Waals heterostructures allows for an easy route to fabricate atomically sharp interfaces with single-atomic-layer precision due to strong covalent bonding within each layer. Fourthly, as can be seen later in this Tutorial Review, one can fabricate electrical contacts to several individual atomic planes within a complex multi-layered structure; this has been a challenging task for the case of adjoining layers grown by MBE.

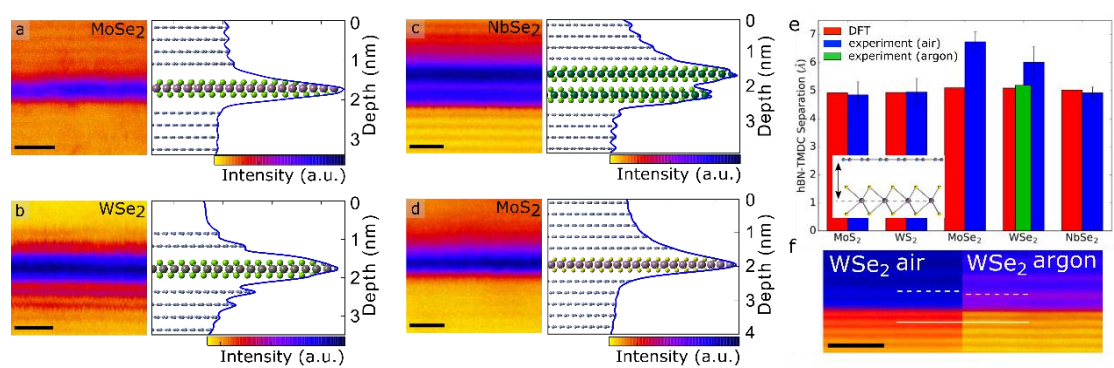

**Figure 8: Cross sectional high-angle annular dark-field (HAADF) scanning transmission electron microscopy (STEM) images of different transition metal dichalcogenides encapsulated in h-BN**. (a) Monolayer MoSe$_2$, (b) monolayer WSe$_2$, (c) bilayer NbSe$_2$, and (d) monolayer MoS$_2$. All scale bars are 1 nm. Each image has an example intensity profile with the y-axis perpendicular to the atomic planes and the x-axis corresponding to the HAADF intensity indicated with colour scale. Underlayed is an atomic model with a scale matching the intensity profile and HAADF

image. The dominant peaks in the intensity profiles correspond to the heavy metals in the transition metal dichalcogenide, and the weaker peaks to the positions of the h-BN planes. (e) Histogram showing the mean interlayer separation values as predicted by density functional theory (DFT) and measured by STEM. DFT and experimental values agree in almost all cases, except for $MoSe_2$ and $WSe_2$ which differ by 1.5 Å and 0.9 Å respectively when stacking is done in air. If stacking is performed in argon (green) experimental values agree with DFT prediction. (f) Shows a difference in separation in a single interface between h-BN and bulk $WSe_2$ as fabricated in air and argon environment. Figure reprinted with permission from Ref.[30] (Copyright 2017 American Chemical Society).

The incredible fact that structures based on h-BN, $WS_2$, $MoS_2$ and graphene have quality comparable to MBE analogues made in ultrahigh vacuum is mainly due to the good fortune of the "self-cleaning" mechanism. Even for stable crystals, adsorbates such as water, atmospheric gases and hydrocarbons cover every surface, unless it is prepared under special conditions. Transmission Electron Microscopy (TEM) studies reveal that graphene is densely covered with hydrocarbons soon after isolation in air, the same applies to other crystals. Fortunately, it turns out that such contamination trapped in between layers is highly mobile and can segregate into isolated 'pockets' (also referred to as 'bubbles' or 'blisters' in the community) leaving large micron-sized clean areas of the heterostructure behind. This limits the homogeneous device size by tens of micrometres for the h-BN/graphene heterostructure and even less for other combinations.

In order to confirm that the segregation process of the adsorbates in the buried interfaces is complete a cross-sectional study was performed by Haigh *et al*.[31] and later expanded to many other material combinations.[30] Figure 8 shows scanning transmission imaging of thin slices extracted from various heterostructures through the use of a focused ion beam; a technique needed to carefully measure the interlayer separation between dissimilar 2D crystals. Indeed, only a handful of materials, such as graphene, h-BN, $MoS_2$, $WS_2$ and $NbSe_2$ have the interlayer spacing expected for a defect-free idealistic interface predicted by density functional theory (DFT). Other materials such as $WSe_2$ and $MoSe_2$ have demonstrated increased interlayer distances, likely caused by the presence of impurities and defects of atomic structure in the interface. While these two materials only demonstrate marginal sensitivity to atmospheric exposure, the majority of layered crystals, currently counting over two hundred species, undergo dramatic surface modification in the air. In bulk crystals deterioration only affects the external layers effectively self-passivating deeper layers, but when exfoliated down to one or two atomic planes, these materials experience deterioration in or complete loss of crystallinity. This has a direct impact on the interface homogeneity even when one of the materials is air-stable. The manifestation of this is nanometer-scale roughness, the absence of any segregated pockets of adsorbates and poor electronic properties of such heterostructures.[27]

In order to address the sensitivity issues, the stacking technique when applied to air-sensitive crystals, is performed in an inert atmosphere, where the 2D materials can be exfoliated, transferred, aligned and encapsulated between air-stable counterparts such as graphene and h-BN (that are completely impermeable to gases and liquids).

Inert atmosphere isolation and the transfer of van der Waals materials have provided a unique platform from which to study 2D materials. This has meant a plethora of previously unexplored 2D materials which in the past degraded before their properties could be measured, are now accessible. Thus, devices produced from black phosphorous remain stable down to a single layer demonstrating reliable field effect transistor.[32] It also allowed the observation of superconductivity in exfoliated monolayers of $NbSe_2$ and FeSe with relatively high $T_c$ of a few Kelvin.[32] Another new material, InSe demonstrated two-dimensional electron gas behaviour with electron mobility's up to 15 000 cm$^2$/Vs, a feature which led to the observation of the quantum Hall effect.[33] More recently, the nanofabrication methods in high-quality inert gas atmospheres have enabled the measurement of mono- and few-layer ferromagnetic 2D materials.[34] We address the reader to section *'6.2. Environmental induced degradation'* where we discuss details of devices fabricated using inert environment technique.

Having discussed constituent layers we now will consider architecture of the devices and development of their complexity. The first van der Waals heterostructure devices were developed to improve the electronic properties and to protect a single "key material" by encapsulating a 2D material between h-BN layers (Figure 9a shows a cross-sectional transmission electron microscopy image of a MoS$_2$ device encapsulated by top and bottom h-BN flakes).[35] As shown in Figure 9b, the mobility of the fully-encapsulated MoS$_2$ devices can be more than one order of magnitude larger than that of unencapsulated devices reaching 34000 cm$^2$/Vs. A similar strategy is used to achieve the high mobility of InSe [33] and BP [32]. In some h-BN-encapsulated materials graphene plays an important role serving as an intermediate contact between a sensitive material and bulk evaporated metallization (see Figure 9a).

A key advantage of van der Waals heterostructures is the additional functionality that arises due to the "third dimension" when there exists more than one active layer in a single stack. An illustrative example of this concept is the assembly of atomically thin p-n junctions, fabricated out of a stack of two different transition metal dichalcogenide monolayers (see Figure 9c) with p- and n-type semiconducting nature.[36,37] The gate traces of the $MoS_2$ and $WSe_2$ layers in Figure 9d demonstrate a marked n- and p-type behaviour respectively and, upon illumination, the device displays the photovoltaic effect (Figure 9e).[36] This device concept was further improved upon by deterministically placing top and bottom graphene electrodes sandwiching the transition metal dichalcogenide bilayer to minimize charge recombination. More recently, this approach has been taken to the extreme by stacking about ten different layers in a band-structure-engineered van der Waals heterostructure that works as a light emitting diode (LED).[38] In this case, the deterministic placement technique allows for the creation of a tailored vertical potential landscape by the fabrication of multiple quantum well heterostructures from layers of transition metal dichalcogenides, h-BN and graphene combined with atomic layer precision. Resulting from the multiple well structures these nanodevices show external quantum efficiencies of up to 8.4%, which compares to the performance of commercially available organic LEDs, and are well-suited for integration in flexible and transparent electronics.

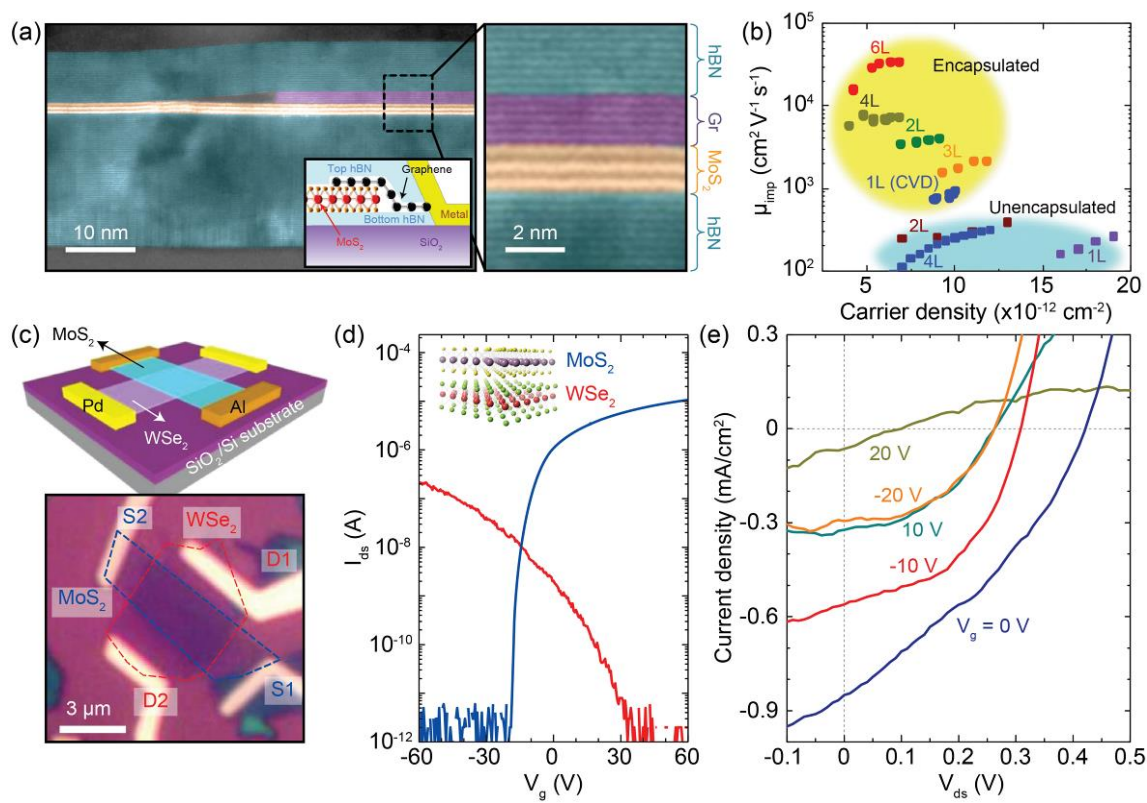

**Figure 9. Fabrication of van der Waals heterostructures.** (a) Cross-sectional STEM image in false-colour of the heterostructure schematically depicted in the inset. The zoom image on the right clearly shows the ultra-sharp interfaces between different layers (graphene, 5L; $MoS_2$, 3L; top h-BN, 8 nm; bottom h-BN, 19 nm). (b) Impurity-limited mobility ($\mu_{imp}$) as a function of the $MoS_2$ carrier density for h-BN encapsulated devices (yellow area) and for unencapsulated $MoS_2$ on $SiO_2$ substrates (blue area). (c) Top: Schematic diagram of a van der Waals-stacked $MoS_2$/$WSe_2$ heterojunction device with lateral metal contacts. Bottom: Optical image of the fabricated device. (d) Gate-dependent transport characteristics at $V_{ds}$ = 0.5 V for individual monolayers of $MoS_2$ (blue curve, measured between S1 and S2) and $WSe_2$ (red curve, measured between D1 and D2). (e) Photoresponse characteristics of the $MoS_2$/$WSe_2$ heterojunction at various gate voltages under white-light illumination. Panels (a) and (b) are adapted from Ref.[35] with permission. Panels (c) to (e) are adapted from Ref. [36] with permission.

## 4. Control of the twisting angle between the stacked layers

A new (nano-device) control knob that has directly emerged from the deterministic placement methods is the twist angle between the atomic/crystalline lattices of the individual layers in a van der Waals stack. This new degree of freedom removes, at least in principle, the symmetry restrictions imposed by the thermodynamic stacking in natural crystals (*vide infra* thermal self-alignment). The superlattices that result from the control of the twist angle host exotic moiré physics that has only just began to be explored and the relevance of this discovery has defined a new field coined by some as 'twistronics'.[39]

The realisation of the significance of the twist angle was triggered by the experimental observation of superlattice Dirac points [40] and the Hofstadter spectrum [11–13] in certain h-BN-encapsulated graphene samples. The latter discovery (the discovery of rotation-dependent moiré physics), was accomplished by transferring a graphene layer on top of h-BN using the van der Waals pick-up method.[9] In addition, the effect was only observed in stacks with near-zero twist angle (< 2º mismatch), when the slight difference in the lattice constants of the graphene and h-BN results in a moiré superlattice with a period of ~14 nm. However, no real control over the twist angle in the fabrication had been achieved so far and the mismatch angle was only inferred *a posteriori* from the transport or scanning probe measurements.[40–42]

Fine control over the twist angle was shown by Ponomarenko *et al.*[12] where artificial bilayers were constructed by sequential transfer of graphene monolayers with a rotational alignment precision of ~1º.

Later, in an attempt to deterministically control the twist angle, Kim *et al*. developed the 'tear-and-stack' technique.[43] This approach is based on the pick-up and transfer technique and consists of selectively picking up a section of a 2D material, rotating it a certain angle and later transferring it on to the section of the same flake left on the substrate. Figure 10 illustrates the process. The implementation of this technique with a high resolution goniometer stage enables a sub-degree control of the twist angle and is instrumental in the fabrication of low-twist-angle heterostructures. Utilising this technique, Cao *et al.*[44]

and Kim et al.[43] reported on h-BN-encapsulated twisted bilayer graphene devices with twist angles equal or smaller than 2º, where insulating states induced by strong interlayer interactions via superlattice modulation were observed. The technique was taken to the current limit by Kim et al.[45] where the sub-degree twist angle regime was explored with a resolution better than 0.2º.

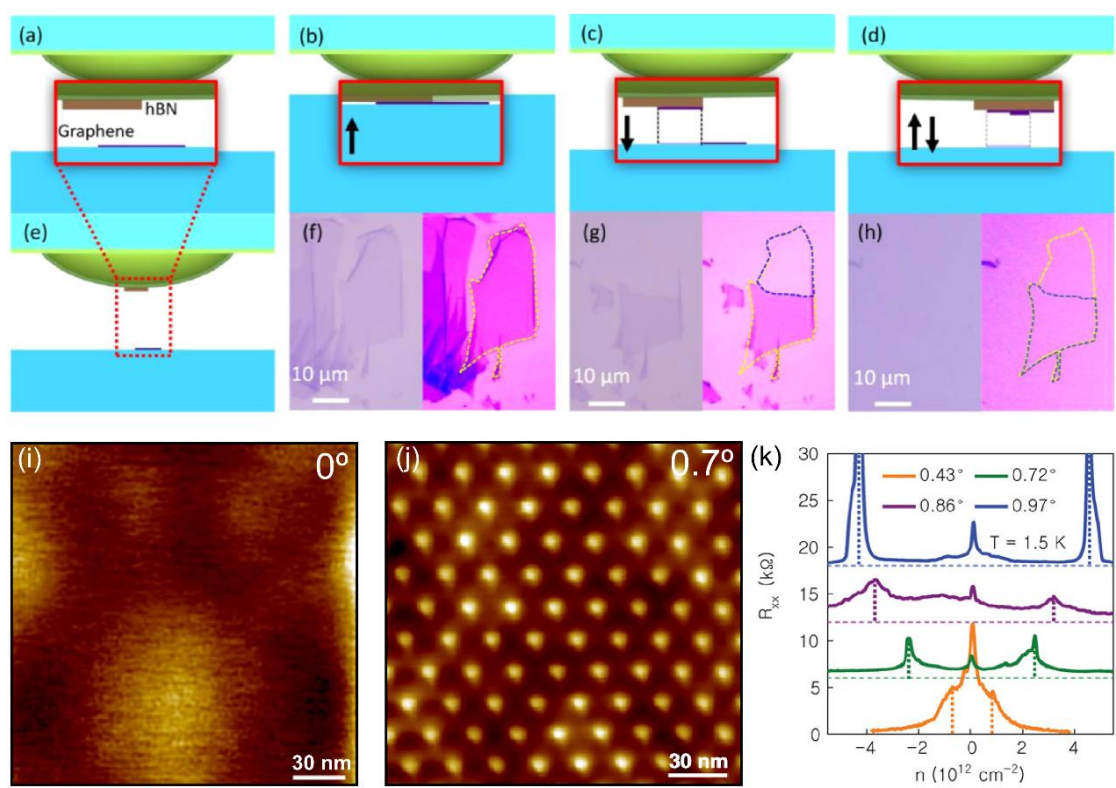

**Figure 10**. **Fabrication of van der Waals heterostructures with well-controlled twist angle.** Illustration of the 'tear-and-stack' transfer technique used to control the rotational alignment in the fabricated artificial bilayers. The method consists on a variation of the van der Waals pick-up method, employing a hemispherical handle substrate. (a) to (e) Schematic of layer pick-up. (f) to (h) Optical images of successive tear and pick-up steps: (b) and (f) the hemispherical handle with a boron nitride flake is in partial contact with a graphene flake (only half of the flake is in contact with the boron nitride), (c) and (g) when the hemisphere is lifted up the graphene flake is teared and part of it is picked-up with the boron nitride, (d) and (h) a second contact with the handle translated laterally or rotated to create an artificial bilayer. Besides each optical image a contrast enhanced image with the graphene contour marked is shown. (i) Scanning tunnelling microscopy image of the topography of an artificial bilayer fabricated with a perfect alignment (note that no moiré pattern is observed). (j) Another example of a sample fabricated with a twist angle of 0.70 ± 0.03º where a triangular moiré pattern with a periodicity of 20.1 nm is observed. (k) Effect of the twisting angle on the electronic properties of the fabricated artificial bilayers: $R_{xx}$ vs. $n$ measured at $T$ = 1.5 K. Panels (a) to (i) were reproduced with permission from Ref.[4]> (Copyright 2016 American Chemical Society). Panels (j) to (k) were reproduced from Ref.[45] with permission (Copyright 2017 National Academy of Sciences).

## 5. Integration of 2D materials into complex devices

In this section, we will describe how deterministic placement of 2D materials can be crucial for their integration into devices with complex architecture and for the fabrication of devices that require a layer-by-layer assembly.

The deterministic placement of 2D materials on arbitrary surfaces with sub-micrometer levels of precision has probably triggered one of the deepest revolutions in the fabrication of nanodevices. These transfer techniques allow both for the integration of selected flakes with other pre-fabricated electronic or optoelectronic nano-devices, and more importantly they also enable the fabrication of complex vertical architectures with atomically-sharp interfaces. The strength of the proximity coupling in these hybrid devices enhances the performance and takes maximum advantage of the quantum properties of 2D materials for technological applications.

An exciting example of 2D materials-based hybrid nanodevices arises from the deterministic transfer of graphene on (micro)optical elements such as (nanometer-scale) waveguides for on-chip optical communications. This type of integration gave rise to high-performance broadband photodetecting modulators based on the evanescent coupling between a graphene flake and a silicon waveguide [14] (see Figure 11a-b). The versatility in the integration of graphene has led to a fast improvement both in responsivity and response time of these waveguide-coupled graphene photodetectors.[15] The devices have quickly approached the benchmarks of the current commercial non-avalanche ones and, more importantly, hold promise for outperforming the current technology soon on account of simple modifications to the device geometry.

2D materials have also been deterministically placed on superconducting microwave cavities in order to couple them optomechanically to microwave photons. In particular, the all-dry transfer technique [10] has been used to fabricate high-Q mechanical resonators with individual graphene flakes that can couple with high-Q superconducting cavities. This type of nanodevices produce a remarkable microwave amplification with a coupling strength that reaches the quantum regime in the motion of graphene.[16,17] (Figure 11c-d).

Deterministic transfer methods also allow one to place 2D materials onto non-conventional substrates where employing conventional nanofabrication techniques would be very challenging. Figure 11e shows few examples of few-layer $MoS_2$ flakes transferred by the all-dry PDMS based deterministic placement method[10] on the surface of an atomic force microscopy cantilever, on the hemispherical lens of a high power LED device and on the core of the free-end of a multimode optical fiber. These examples illustrate the power of deterministic placement methods to integrate 2D materials in applications that are not compatible with the standard approach (exfoliation and bottom up fabrication).

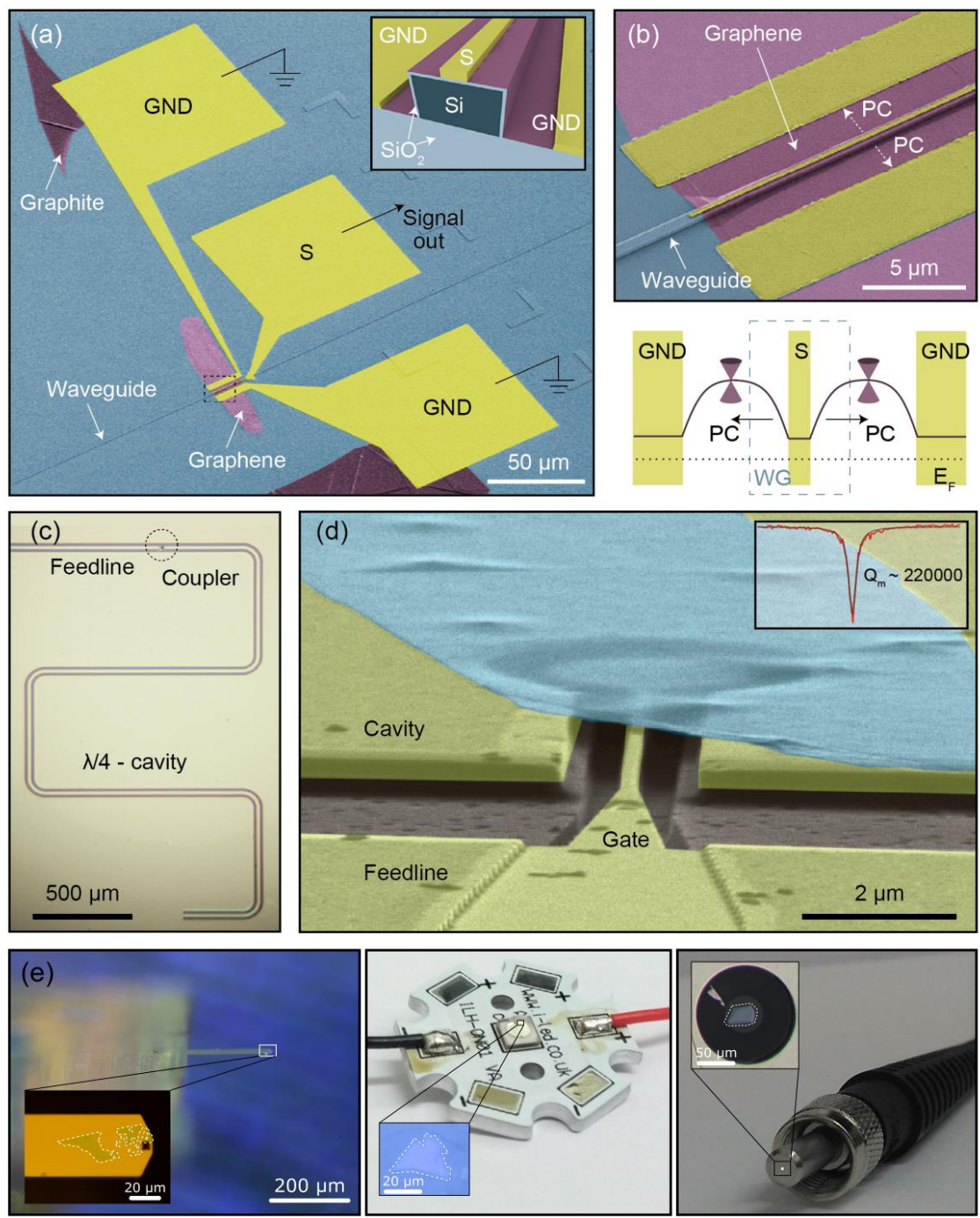

**Figure 11**. **Integration of 2D materials into complex devices.** A waveguide-integrated graphene photodetector and a multilayer graphene mechanical resonator coupled to a superconducting microwave cavity. (a) Coloured scanning electron micrograph of a waveguide-integrated graphene photodetector. The 'active region' of the graphene sheet is shown in violet. The inset shows a cross-section of the device. (b) Top: Enlarged view of the section highlighted by the black dashed rectangle in a. Bottom: Schematic of the device, PC stays for photocurrent. (c). Optical image of a superconducting cavity in a quarter-wave coplanar waveguide geometry. Microwave photons are coupled in and out of the cavity via a feedline that is capacitively coupled to the coplanar waveguide center conductor by a gate capacitor (coupler). (d) Coloured tilted angle scanning electron micrograph showing a 4-μm diameter multilayer (10-nm-thick) graphene resonator (cyan) suspended 150 nm above the gate. Inset: Quality factor of the graphene mechanical resonator. Panels (a) and (b) were reproduced from Ref.[15] with permission. Panels (c) and (d) were reproduced from Ref.[16] with permission. (e) Examples of deterministic placement of few-layer $MoS_2$ onto non-conventional substrates by the all-dry PDMS deterministic transfer method[10]: (left) on an atomic force microscopy cantilever, (center) on the hemispherical lens of a LED and (right) on the core of a multimode fiber optic. Panel (e) has been provided by the authors.

## 6. Challenges and outline

Here we will discuss about the challenges that remain to be solved. We will focus on two main issues: the presence of interlayer adsorbates and the environmental instability of certain 2D materials. The ongoing efforts to tackle these problems will be introduced as well.

### 6.1. Interlayer contaminants

The presence of interlayer contaminants is an important issue in the fabrication of van der Waals heterostructures since it can lower the performances of the 2D materials stack and limit realistic homogeneous device size. As it was mentioned previously, trapped contamination segregates into isolated bubbles typically tens to hundreds of nanometres in size, driven by equilibrium between the cost of elastic energy to form such bubbles and the gain in the surface van der Waals energy from the formation of clean interface regions. Figure 12a shows AFM topography of a typical heterostructure fabricated by stamping graphene on top of h-BN. From the topographic AFM image wrinkles and bubbles can be seen in the top graphene layer pinned to an atomic terrace in h-BN substrate. If the bubble is broken using a diamond AFM tip the contamination is able to escape behaving as a viscous liquid until the bubble is resealed. However when a plasma or ion etching processes are used during the device fabrication such bubbles demonstrate a considerable etch time similar to organic polymers and behave more like a solid due to crosslinking processes. The morphology and chemical composition of the contamination was studied in Ref.[31] revealing that bubbles indeed consist of dense amorphous material based on carbon and oxygen. Figure 12c illustrates the effect of thermal annealing on the bubbles present in a graphene flake deposited on top of an h-BN flake by showing AFM images of the stack before and after annealing at 200 °C and 500 °C. After annealing the stack in Ar/$H_2$ gas at 200 °C, the smaller bubbles aggregate into larger bubbles in order to reduce the total surface energy (in the process known as Ostwald ripening) as can be seen by comparing the line-profiles in the figure. The authors reported that the number of bubbles decreased after annealing, while the size and height of bubbles increased. Increasing the annealing temperature to 500 °C led to the rupture of bubbles and to the escape of part of the adsorbates as shown in the rightmost panel of Figure 12. The surface of graphene after the annealing has larger clean and flat areas compared to the initial surface, leading to better performances of the heterostructure such as an increased mobility.

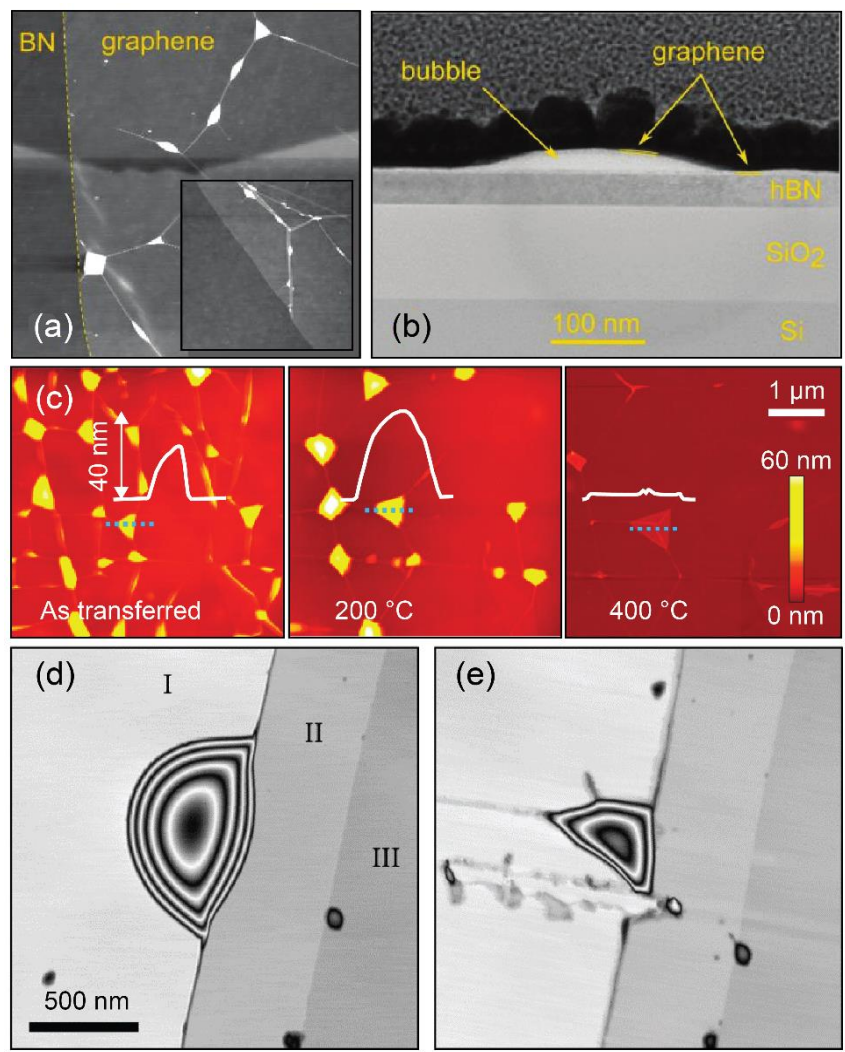

**Figure 12. Presence of interlayer contaminants in the as-transferred samples and cleaning process.** (a) AFM topography image of graphene transferred onto h-BN, showing the presence of bubbles where the hydrocarbon contamination is aggregated. Scan area 15 μm × 15 μm (the inset is a zoomed in image with an area of 1.5 μm × 1.5 μm). (b) Cross-sectional TEM image of graphene on h-BN illustrating the hydrocarbon contamination present in the bubbles. (c) AFM images of graphene transferred to h-BN before and after annealing at different temperatures. The annealing process allows the interlayer adsorbates to diffuse and to coalesce into larger bubbles. At 500 ºC the bubbles burst and part of the adsorbates escape. (d) and (e) show AFM images of graphene transferred to h-BN with a bubble before and after scratching it with an AFM tip. The contamination contained in the bubble leaks off after bursting the bubble with the AFM tip but the van der Waals interaction re-seals the bubble. Panels (a) and (b) were reproduced from Ref.[27] with permission. Panel (c) has been adapted from Ref.[20] with permission of the authors. Panel (d) and (e) were reproduced from Ref.[31] with permission.

While such contamination presents a global problem for the entire field there are strategies that help to maximize clean areas of heterostructures. These involve precise control over the transfer mechanics at the moment when two flakes first engage in contact.

This process of initial lamination should be made very slowly so that when the contact line is progressing across the device gradually more contamination can escape the closing interface and avoid being trapped. This can be further improved by heating the sample to enhance the diffusion of contaminants.[29]

### 6.2. Environmental induced degradation

The control of the environmental conditions during the transfer can be relevant not only to avoid the presence of interlayer contaminants, as discussed above, but to work with materials that tend to degrade upon exposure to air. Exfoliated flakes of black phosphorus, for example, are highly hygroscopic and tend to uptake moisture from air (see Figure 13)[46] and its exposure to this condensed water on the surface seems to degrade its mobility [46,47] and increases its sheet resistance.[48] Another example of an air-sensitive 2D material is niobium diselenide, a type-II s-wave superconductor with a transition temperature, $T_C$, of 7.2 K. Recent works found that thin $NbSe_2$ tends to oxidize hampering the fabrication of ultrathin devices.[32,49]

The pristine electronic properties of black phosphorus or $NbSe_2$ can be preserved by encapsulating them between two flakes of h-BN under oxygen- and moisture-free conditions.[32,5>] These devices show superconducting transition in monolayer $NbSe_2$ and very high carrier mobility for monolayer BP. These encapsulated devices also show no evidence of degradation and photo-oxidation, even after air exposure for many weeks.[32]

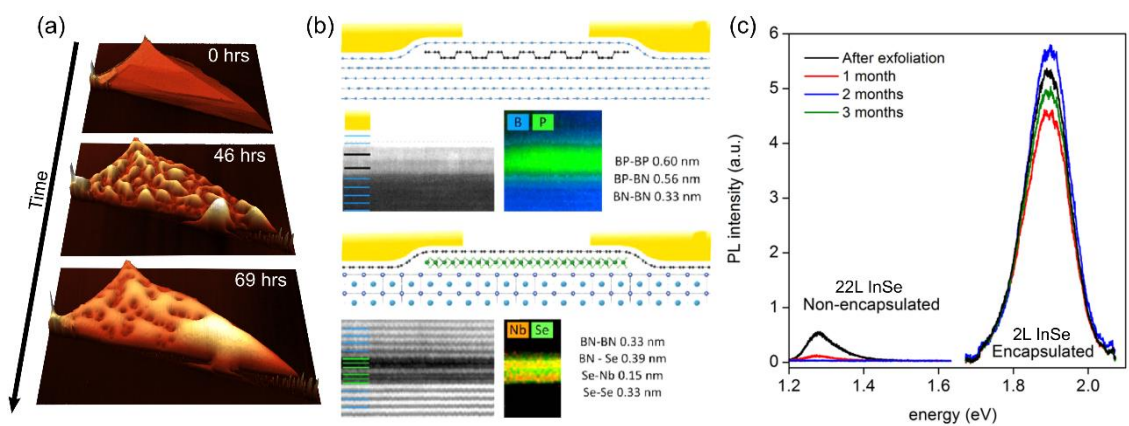

**Figure 13**. **Environmental induced degradation of 2D materials and their encapsulation.** Atomic force microscopy topography images acquired on a non-encapsulated black phosphorus flake as a function of time. Due to the high hygroscopic character of black phosphorus it uptakes moisture from air. (b) Schematics of the glovebox encapsulated BP and NbSe$_2$ devices and Cross-sectional high-angle annular dark-field STEM images and STEM images superimposed with elemental profiles. (c) Evolution in time of the photoluminescence response of encapsulated (2L) and non-encapsulated (22L) InSe flakes (excitation: $E$=2.33 eV, $P$=1.35 mW). While the encapsulated flake remains stable over months, the non-encapsulated rapidly degrades upon exposure to environmental conditions. Panel (a) has been adapted from Ref.[4]> with permission of the authors. Panel (b) has been reproduced from Ref.[32] with permission. Panel (c) has been reproduced from Ref. [33] with permission.

Another example of material that could be studied thanks to the development of the exfoliation and encapsulation process within a glovebox with an inert atmosphere is the indium monoselenide.[33] This material degrades upon exposure to air but it remains stable over months once encapsulated between h-BN flakes (see Figure 13c). InSe demonstrated electron mobilities exceeding 15000 cm$^2$V$^{-1}$ s$^{-1}$ at liquid helium temperatures, allowing for the observation of the fully developed quantum Hall effect and making it currently the third best two-dimensional electron gas (2DEG) material for electronic applications. Photoluminescence spectroscopy reveals that the bandgap increases by more than 0.5 eV with decreasing the thickness from bulk to bilayer InSe. The band-edge optical response vanishes in monolayer InSe, which is attributed to the monolayer's mirror-plane symmetry.

Despite the recent progress on 2D materials beyond graphene with high electronic quality reported to date (MoS$_2$, WSe$_2$, BP and InSe), they show their peak performance for devices with thicknesses ranging between 5 and 10 layers. When thinner devices, such as single and bilayer FETs are fabricated their electronic properties such as carrier mobility, residual doping and homogeneity drastically deteriorate by at least two orders of magnitude. One possible reason for this may be residual water and oxygen in the environmental transfer chamber causing surface damage more pronounced in thinner crystals (state of the art glovebox chambers only reach $10^{-4}$ mbar partial oxygen and water pressure).

Another reason for poorer performance of thin crystals can be significantly less efficient "self-cleaning" process as compared to few-layer films of the same material. Such incomplete segregation of the adsorbates is likely causing scattering and rough interface morphology with complex local strain. Unfortunately using argon or nitrogen environments, even with the best quality commercially available gas purification systems, does not offer a remedy to the "bubble" problem and only partially reduces the amount of hydrocarbon contamination. While current environmental fabrication setups make many new materials accessible for 2D device fabrication, it is also clear that much further work is needed to develop new fabrication strategies to achieve cleaner van der Waals interfaces.

## 7. Conclusions

In recent years we have witnessed the development of ingenious deterministic transfer techniques that bring one step closer Feynman's visions of a future where it is possible to control the layered structure of materials, arranging the atoms in the way we want them. These deterministic placement methods have made it possible to fabricate artificial materials, not existing in nature, that showed new physical phenomena and to assemble complex devices that would be otherwise almost impossible to fabricate by conventional bottom-up nanofabrication approaches. We have shown how deterministic placement methods also open the door to fabricate twisted structures with an exquisite control over the relative orientation of the lattices forming the stack. This has made it possible to probe novel physics by fabricating devices previously inaccessible in conventional MBE grown heterostructures. In this Tutorial Review we have revised the recent progress on deterministic placement of 2D materials, describing and comparing the different alternative methods available in the literature. We also illustrate their potential to fabricate heterostructures by artificial stacking of dissimilar 2D materials, to assemble complex devices and to fabricate artificial bilayer structures where the layers present a user-defined rotational twisting angle. In summary, these deterministic placement techniques offer unprecedented control over the fabrication of samples based on 2D materials and open up exiting research avenues for the near future.

## ACKNOWLEDGEMENTS

The authors are grateful to Matthew Hamer for a critical reading of our manuscript. AC-G acknowledges funding from the European Commission under the Graphene Flagship, contract CNECTICT-604391. RF acknowledges support from the Netherlands Organisation for Scientific Research (NWO) through the research program Rubicon with project number 680-50-1515. DPdL acknowledges support from MINECO through the program FIS2015-67367-C2-1-p. PJH acknowledges support by the Center for Integrated Quantum Materials under NSF grant DMR-1231319, the Gordon and Betty Moore Foundation's EPiQS Initiative through Grant GBMF4541, the US DOE, BES Office, Division of Materials Sciences and Engineering under Award DE-SC0001819, and NSF award DMR-1405221. Device fabrication has been partly supported by the Center for Excitonics, an Energy Frontier Research Center funded by the US Department of Energy